\documentclass[structabstract]{aa}
\usepackage{natbib}
 \bibpunct{(}{)}{;}{a}{}{,}

\usepackage{graphicx}

\usepackage{txfonts}
\def\kms{km s$^{-1}$\space}
\def\kmsno{km s$^{-1}$}
\def\micron{$\mu$m\space}

\def\micronno{$\mu$m}
\def\arcsecno{$^{\prime\prime}$}
\def\arcsec{$^{\prime\prime}$\space}

\def\arcmin{$^{\prime}$\space}
\def\deg{$^{\circ}$\space}

\def\degno{$^{\circ}$}
\def\h2{H$_2$}

\def\cii{[C\,{\sc ii}]\space}

\def\ciino{[C\,{\sc ii}]}

\def\hi{H\,{\sc i}\space}
\def\hino{H\,{\sc i}}
\def\hii{H\,{\sc ii}\space}

\def\12co{$^{12}$CO}
\def\13co{$^{13}$CO}
\def\c18o{C$^{18}$O}

\def\nii{[N\,{\sc ii}]\space}
\def\niino{[N\,{\sc ii}]}

\def\C+{C$^+$}
\def\h2{H$_2$}
\def\cm3{cm$^{-3}$}
\def\cm3s{cm$^{-3}$\space}
\def\cm2{cm$^{-2}$}
\def\cm2s{cm$^{-2}$\space}

\begin{document}
%%%%%%%%%%%%%%%%%%%%%%%%

%   \title{Ionized gas at the Scutum tangency traced in \nii and \cii}%\thanks{The}
    \title{Ionized gas in the Scutum spiral arm  as traced in \nii and \cii}%\thanks{The}

\titlerunning{Ionized gas in the Scutum spiral arm as traced in \nii and \cii}

%\authorrunning{Langer, Velusamy, Requena-Torres, Wiesemeyer, ???}
\authorrunning{W. D. Langer et al.}

   \author{W. D. Langer
             \inst{1},
 	 T. Velusamy
                  \inst{1},
                  P. F. Goldsmith
                  \inst{1}, 	
          J. L. Pineda
	           \inst{1},
           E. T. Chambers
          	 \inst{2},
	 G. Sandell
		 \inst{2},
	           C. Risacher
            	 \inst{3},
	           \and
	  K. Jacobs
	    \inst{4}
 %	              \and
		 }

         % \offprints{W.\,D.\,Langer \email{William.Langer@jpl.nasa.gov}}

   \institute{Jet Propulsion Laboratory, California Institute of Technology,
              4800 Oak Grove Drive, Pasadena, CA 91109, USA\\
              \email{William.Langer@jpl.nasa.gov}
		  \and		               
	 SOFIA-USRA, NASA Ames Research Center, MS 232-12, Moffett Field, CA 94035-0001, USA
 		\and		
             Max-Planck-Institut f$\ddot{\rm u}$r Radioastronomie,
		Auf dem H$\ddot{\rm u}$gel 69, 53121 Bonn, Germany
		\and	
                  I. Physikalisches Institut der Universit$\ddot{\rm a}$t zu K$\ddot{\rm o}$ln, Z$\ddot{\rm u}$lpicher Strasse 77, 50937,
                  K$\ddot{\rm o}$ln, Germany                	 
%             \thanks{Who do we thank?}
             }

   \date{Received 18 May, 2017; Accepted 26 July, 2017}
%\abstract{}{}{}{}{}
% 5 {} token are mandatory

%__________________________________________________________________

% \abstract
  %context heading (optional)
%{We}

\abstract
{The warm ionized medium (WIM) occupies a significant fraction of the Galactic disk. Determining the WIM properties at the leading edge of spiral arms is important for understanding its dynamics and cloud formation.}
 {To derive the properties of the WIM at the inner edge of the Scutum arm tangency, which is a unique location in which to disentangle the WIM from other components, using the ionized gas tracers C$^+$ and N$^+$. }
 {We use high spectral resolution \cii 158 \micron  and \nii 205 \micron fine structure line observations taken with the upGREAT and GREAT instruments, respectively, on SOFIA, along with auxiliary  \hi and $^{13}$CO observations.  The observations consist of samples in and out of the Galactic plane  along 18 lines of sight (LOS) between longitude 30\deg and 32\degno. }
 { We detect strong \nii emission throughout the Scutum tangency. At $V_{LSR}$ = 110 to 125 \kms where there is little, if any, $^{13}$CO, we are able to disentangle the \nii and \cii emission that arises from the WIM at the arm's inner edge.   We find an average electron density $\sim$0.9 cm$^{-3}$ in the plane, and $\sim$0.4 cm$^{-3}$ just above the plane.  The \nii emission decreases exponentially with latitude with a scale height $\sim$55 pc. For $V_{LSR}<$110 \kms there is \nii emission tracing highly ionized gas throughout the arm's molecular layer.  This  ionized gas has a high density, $n$(e) $\sim$30 cm$^{-3}$, and a few percent filling factor. We also find evidence for \cii absorption by foreground gas.}
{\nii and \cii observations at the Scutum arm tangency reveal a highly ionized gas with average electron density about 10 to 20 times those of the interarm WIM, and is best explained by  a model in which the interarm WIM is compressed as it falls into the potential well of the arm. The widespread distribution of \nii in the molecular layers  shows that high density ionized gas is distributed throughout the Scutum arm.  The electron densities derived from \nii for these  molecular cloud regions are $\sim$30 cm$^{-3}$, and probably arise in the ionized boundary layers of clouds.  The \nii detected in the molecular portion of the spiral arm arises from several cloud components with a combined total depth  $\sim$8 pc. This \nii emission most likely arises from   ionized boundary layers, probably the result of the shock compression of the WIM as it impacts the arm's neutral gas, as well as from extended \hii regions.}

{} 

\maketitle

%________________________________________________________________

%%%%%% SECTION 1 INTRODUCTION %%%%%%%%%%%%%%%%%%%%%%%%%
%%%%%% SECTION 1 INTRODUCTION %%%%%%%%%%%%%%%%%%%%%%%%%

\section{Introduction}
\label{sec:introduction}

\begin{figure}
% \begin{figure*}[!ht]
 \centering
             \includegraphics[width=9cm]{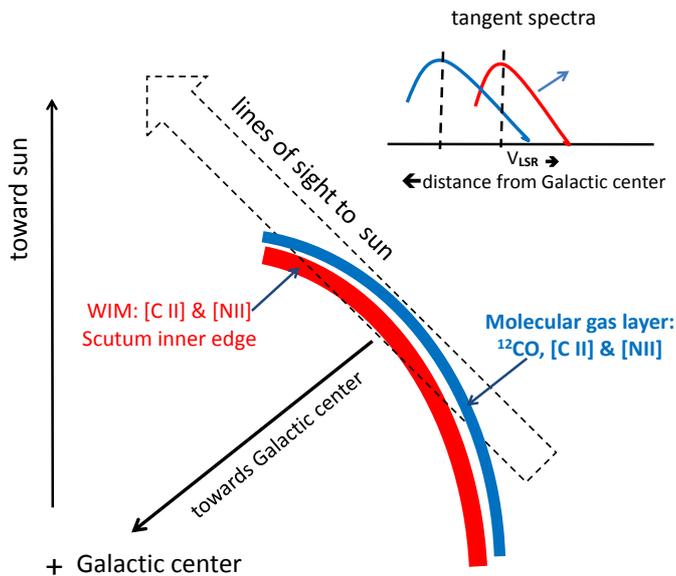}
      \caption{Schematic view of the Scutum spiral arm tangency showing the structure of the gas layers (adapted from Figure 6 in \cite{Velusamy2015}. The emission of key gas tracers along the tangency, \ciino, \niino, and CO help distinguish the WIM and molecular gas layers indicated as red and blue colors, respectively. Note the long path length along the tangency which improves the sensitivity to the weak \cii and \nii emission from the WIM. A schematic of the velocity ($V_{LSR}$) profile (shown in the inset) of the corresponding spectral line intensities for each layer, demonstrates why it is possible to separate the emission from the WIM from the neutral gas components. Note that this cartoon is intended to be a schematic and is not to scale.} 
         \label{fig:fig1}
 \end{figure}

The interstellar medium (ISM) is a dynamic environment in which gas cycles from a diffuse ionized state to dense star forming molecular clouds and back.  Stars and supernovae provide energy to disrupt and ionize the gas and generate dynamical flows within and above the disk.  Spiral density waves at the leading edge of the arm compress the ionized interarm gas and initiate one of the processes of cloud formation.  The dynamics of the arm and the energy sources within the arm determine the distribution of different interstellar gas components within and above the disk.  The study of the spiral density waves is thus important for understanding how clouds form and galaxies evolve.  The spiral tangent regions are ideal laboratories to study the interaction of the interstellar gas and spiral density waves in the Milky Way as they provide a unique viewing geometry with large enough path lengths to detect the diffuse (atomic or ionized) spiral arm components.  This unique viewing geometry of a spiral arm is illustrated in Figure~\ref{fig:fig1} for the Scutum tangency (adapted from Figure 6 in \cite{Velusamy2015}) in which the different gas layers, ionized, atomic, and molecular are shown along with their emission line profiles as a function of $V_{LSR}$. These tangencies are located at $l$ $\sim$ 31\degno, 51\degno, 284\degno, 310\degno, 328\degno, and 339\deg \citep{Vallee2008a}, although there is some scatter, $\sim$4\degno, in the identification of the location of the tangency depending on which tracer is used.

The atomic and molecular components of the ISM in the spiral arms are well traced in emission by \hi and CO, respectively, but the photon dominated regions (PDR), CO-dark \h2 and the ionized gas has been less well studied in emission owing to the difficulty of observing key tracers from the ground.  The fine-structure transition of ionized carbon, \ciino, at 158 \micronno, traces almost all the warm (T$>$ 35K) portions of the ionized ISM. However, \cii alone cannot distinguish fully ionized hydrogen gas from weakly ionized gas (carbon ionized but hydrogen neutral).  In contrast, nitrogen, with its 14.53 eV ionization potential, has fine-structure lines, \niino, at 122 \micron and 205 \micron that arise only in highly ionized regions. Here we investigate the influence of the spiral arms on the distribution and dynamics of the highly ionized gas in and out of the plane by observing velocity resolved \nii 205 \micron and \cii158 \micron emission in the Scutum spiral arm tangency ($l \sim$ 31\degno), taken with the German Receiver for Astronomy at  Terahertz Frequencies \cite[GREAT\footnote{GREAT and upGREAT are a development by the MPI  f$\ddot{\rm u}$r Radioastronomie and the KOSMA/Universit$\ddot{\rm a}$t zu K$\ddot{\rm o}$ln, in cooperation with the MPI  f$\ddot{\rm u}$r Sonnensystemforschung and the DLR Institut  f$\ddot{\rm u}$r Planetenforschung.};][]{Heyminck2012} and the upGREAT$^1$ array \citep{Risacher2016}, respectively, onboard the NASA/DLR Stratospheric Observatory for Infrared Astronomy \cite[SOFIA;][]{Young2012}.

The large-scale structure of spiral arms in the Milky Way is a subject of great interest for understanding the dynamics of the Galaxy and for interpreting its properties.  Modeling the Galactic spiral structure is based in part on data at the spiral arm tangents in different gas and stellar tracers.  However, each of these tracers can occupy a separate lane across the arm reflecting the evolution of gas from low- to high-density clouds as they are swept into the arm's gravitational potential.  Using Herschel HIFI \cii maps along with \hi and CO maps,  \cite{Velusamy2012,Velusamy2015} showed that the gas in the Scutum, Crux, Norma, and Perseus arms was arranged in layers and revealed an evolutionary transition from lowest to highest density states.  In particular the geometry of the tangencies made possible the detection of the low density WIM in \cii and revealed a higher density WIM at the inner edge of the arm than in the interarm gas. They suggested this increase was a result of the compression of the WIM at the leading edge.  Here we use both \cii and \nii observations of the Scutum arm tangency at $l \sim$31\deg to study the interaction of the spiral arm potential with the ionized interarm gas. 

% Figure 1
% Figure 1

This paper is organized as follows.  In Section 2 we present the observations and data reduction, while in Section 3 we describe the distribution of \cii and \nii in the Scutum tangency.  In Section 4 we derive  the properties of the ionized gas, including density and scale height, using \cii and \niino, and discuss the possible mechanisms responsible for the distribution of \niino. Section 5 summarizes the results.

%__________________________________________________________________

%%%%%% SECTION 2 OBSERVATIONS %%%%%%%%%%%%%%%%%%%%%%%%%
%%%%%% SECTION 2 OBSERVATIONS %%%%%%%%%%%%%%%%%%%%%%%%%

\section{Observations }
\label{sec:observations}

\begin{figure}
% \begin{figure*}[!ht]
 \centering
             \includegraphics[angle=-90,width=9cm]{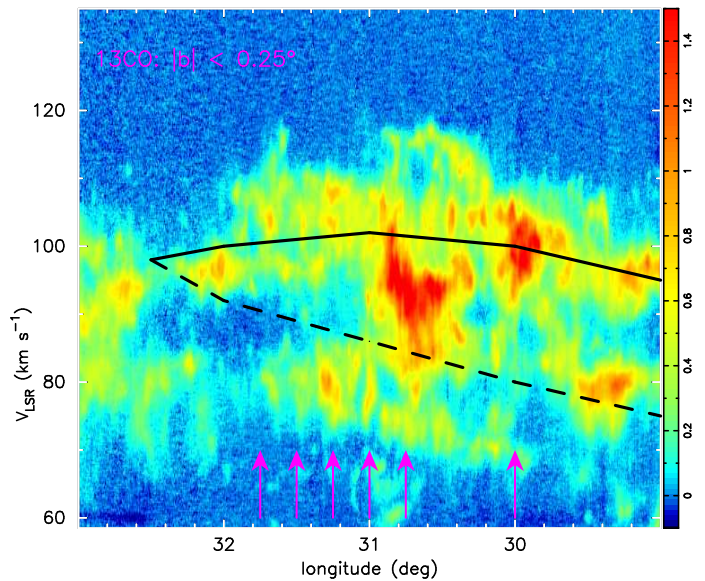}
      \caption{A \13co longitude--velocity plot of the Scutum arm in the region $l \sim$29\deg to 33\deg at $b$=0\degno, where we mark the inner (black solid line) and outer (dashed black line) tangencies taken from \cite{Reid2016} using their fit to the near--far distances.  The arrows at the bottom indicate Galactic longitudes of the lines of sight observed in \cii and \nii with upGREAT and GREAT, respectively. }
         \label{fig:fig2}
 \end{figure}

The Scutum arm is located about 4 kpc from the Galactic Center and wraps more than half way around the Galaxy.  In Figure~\ref{fig:fig2} we show a \13co  longitude--velocity map of the tangency near $l \sim$31\deg and have marked the inner and outer tangencies  derived from \cite{Reid2016} with black lines.  The \13co traces the molecular clouds in this region.   The $^{13}$CO ($l$--$V_{LSR}$) map at Galactic latitude, $b$ = 0\deg is derived from the Galactic Ring Survey (GRS) data \citep{Jackson2006}. 

% Figure 2
% Figure 2

To compare the spatial and velocity structure of the spiral arm gas components in both the molecular (neutral) and the ionized gases in the Scutum arm we made small--scale cross maps of \cii and \nii in longitude along and in latitude above a spiral arm tangency with the GREAT single pixel receiver for \nii and the upGREAT 7-pixel array for \ciino. The  \nii and \cii emission across the arm allows us to examine the impact of the spiral arm potential on the ionized gas components.   To examine the spatial and velocity structures of these gas components across the tangency and perpendicular to the Galactic plane we observed \nii and \cii along 18 lines of sight (LOS) across the Scutum tangency from 30\fdg0 to 31\fdg75 covering $b$ from -0\fdg25 to 1\fdg7.  These 18 LOS are indicated in Figure~\ref{fig:fig2} by crosses superimposed on a 21 cm continuum map \citep{Stil2006} of the Scutum tangency region. We have also indicated the primary, secondary, and tertiary reference sky positions used in the observations (see discussion below).  

% Figure 3
% Figure 3
\begin{figure}
% \begin{figure*}[!ht]
 \centering
             \includegraphics[width=8.5cm]{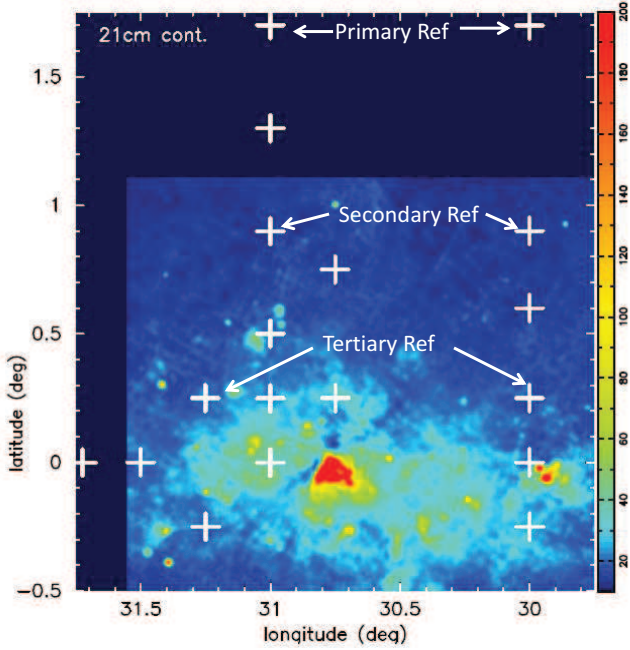}
      \caption{The 18 lines of sight observed in \cii and \nii are indicated by white crosses superimposed on a 21 cm continuum map of the Scutum tangency region \citep{Stil2006}. The primary reference off positions are labeled as are the secondary and tertiary reference positions.}
         \label{fig:fig3}
 \end{figure}

We observed the Scutum tangency in the ionized carbon (C$^+$) $^2$P$_{3/2}$ -- $^2$P$_{1/2}$  fine structure line, \ciino, at 1900.5369 GHz ($\lambda  \sim$157.741 \micronno), using upGREAT \citep{Risacher2016} and the ionized nitrogen (N$^+$) $^3$P$_1$--$^3$P$_0$ fine structure line, \niino, at 1461.1338 GHz $(\lambda \sim$ 205.178 \micronno) using GREAT \citep{Heyminck2012} onboard SOFIA \citep{Young2012}.  The upper state of C$^+$, $^2$P$_{3/2}$, lies at an energy, $E$/k, about 91.2 K above the ground state, and the upper states of N$^+$, $^3$P$_1$ and $^3$P$_2$, lie at 70.1 K and 188.2 K, respectively, above the ground state.  Our program (proposal ID 04\_0033; PI Langer) was  part of the Guest Observer Cycle 4 campaign.  The observations were made between May 12 and June 10, 2016 on six flights spread over SOFIA flights \#296 to \#308. Both \cii and \nii spectra  were observed simultaneously in GREAT configuration: LFA+H+V/L1 with \cii in LFA:LSB and \nii in L1:USB.  The typical observing time on each target was in the range of 15 to 30 minutes.
 
\cii and \nii are widespread throughout the Galaxy so finding a clean off position for proper calibration is a challenge in observing the Scutum spiral arm.  The telescope on SOFIA cannot slew more than 0\fdg4 to 0\fdg8 from the ON to the OFF position, and it was thus necessary to observe \cii and \nii in a series of steps in $b$ starting at a latitude where the emission is weak enough to provide a clean OFF position.  In practice even the largest $b$ value was not always absolutely clean of emission, but it was generally weak enough to use as a reference position. The observing scenario, therefore, used a set of primary, secondary and tertiary reference positions located at $b$ = 1\fdg7, 0\fdg9, and 0\fdg25, respectively as shown in Figure~\ref{fig:fig3}.  The primary reference positions, the LOSs at $b$=1\fdg7, were observed with respect to an OFF position at $b$=2\fdg1 at the same longitude.  These LOS at ($l$,$b$) = (30\fdg0,1\fdg7) and (31\fdg0,1\fdg7) were then used as primary reference positions for observations at lower values of $b$, two of which at $b$= 0\fdg9 then could be used as secondary reference positions (see Figure~\ref{fig:fig3}) for LOS at yet  lower values of $b$.  This procedure was repeated  one more time and a set of tertiary reference positions were established at $b$ = 0\fdg75 (see Figure~\ref{fig:fig3}). In this scheme each target LOS used the nearest (up to within 0\fdg8) primary/secondary/tertiary reference position.  The calibrated emission spectra at the reference positions were than added incrementally in a hierarchical manner to the respective target spectra.  

\begin{figure}
% \begin{figure*}[!ht]
 \centering
                  \includegraphics[width=5.9cm,angle=-90]{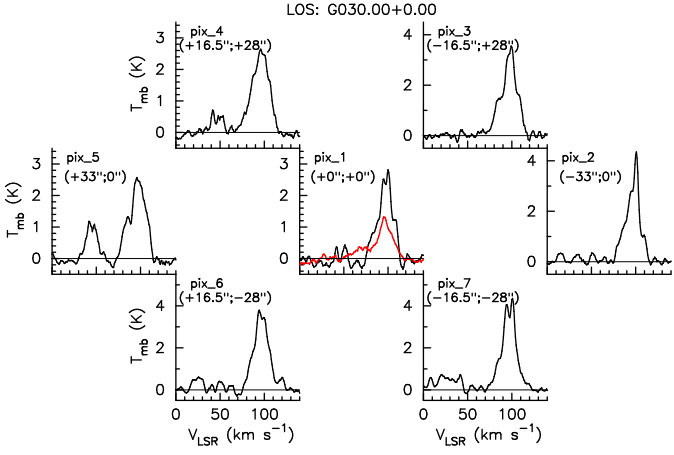}
      \caption{The seven \cii spectra observed by upGREAT (black) and the central \nii spectrum (red) observed by GREAT towards ($l$,$b$)= (30\fdg0,0\fdg0).  While the overall shapes of the \cii lines across the Scutum spiral arm ($V_{lsr}$ = 60 to 130 \kmsno) within each pixel are similar, the small differences among them  demonstrate that there is structure due to individual cloud components. The \nii emission in the central pixel  is detected across the entire velocity range of \cii in the Scutum arm.}
         \label{fig:fig4}
 \end{figure}

 The dual-polarization 7-pixel upGREAT array has a FWHM beam size  of 14\arcsec at 1.9 THz. The array  is arranged in a hexagonal pattern with a central beam.  The beam spacing is approximately two beam widths and the array has a footprint about 67\arcsec across. At the distance to the Scutum arm tangency, $\sim$ 7 kpc (see Section~\ref{sec:discussion}) , each pixel corresponds to $\sim$0.5 pc and the array stretches across  $\sim$2.3 pc. The \nii lines were observed with the GREAT receiver which has only one dual-polarization pixel with a FWHM beam size of 19\arcsec at 1.4 THz. The \nii single pixel is centered on the central \cii (pix\_1), as shown in Figure~\ref{fig:fig4} for (30\fdg0,0\fdg0).  It can be seen that to first order the overall shape of the \cii emission  is similar across the array, however there are small scale variations reflecting the individual gas components.  This variation on 20\arcsec scale is not surprising considering that the 7-pixel footprint of upGREAT covers $\sim$2.3 pc across, and probably includes a number of different cloud components. Only for the central pixel (pix\_1) do both \cii and \nii have identical pointings.  Therefore for all analysis comparing their intensities we use only the \cii central pixel.  For all other analysis of large scale or global characteristics of \cii emission, for example in comparison to molecular gas traced by $^{13}$CO, we use the \cii spectra averaged over all 7 pixels because this average roughly matches the beam sizes of the CO auxiliary data.  The intensities have been converted to main beam temperature, T$_{mb}$(K), using beam efficiencies, $\eta_{mb}$(\ciino)=  0.65 and $\eta_{mb}$(\niino)=  0.66 with a forward efficiency, $\eta_{for}$=  0.97   \citep{Rollig2016}.   The rms noise for  T$_{mb}$(K) in  the calibrated spectra shown in Figure~\ref{fig:fig5}  are in the range of 0.05 to 0.17 K for \niino, 0.04 to 0.25 K for \cii averaged over 7 pixels, and 0.08 to 0.30 K in pixel \#1.  To avoid the propagation of noise from the off position, we add to the target spectrum only off source spectral intensities that are above 1.5 times the rms noise.  This approach is particularly important to minimize adding noise to the spectra since a majority of the target LOS spectra use multiple off source spectra for reference. The large variation in the noise levels in these spectra are due to the differences in the  observing duration used for each LOS.

% Figure 4
% Figure 4
% \end{figure*}

In addition to the SOFIA \cii and \nii observations we use auxiliary \hi and $^{13}$CO(1-0) data from public archives,  \nii from {\it Herschel} PACS and HIFI at (30\fdg0,0\fdg0) and  (31\fdg2766,0\fdg0) \citep{Goldsmith2015,Langer2016}, and \cii with HIFI \citep{Langer2016}. The   \hi 21 cm data are taken from the VLA Galactic plane survey (VGPS)  \citep[][]{Stil2006}.  The $^{13}$CO(1-0) data from the Galactic Ring Survey (GRS) \citep{Jackson2006}. We extracted the spectra for the LOSs observed by SOFIA from these survey data, which  are available as Galactic longitude--latitude--V$_{LSR}$ spectral line data cubes. Both VGPS and GRS surveys have comparable angular resolution   of 1\arcmin and 40\arcsecno, respectively.  The spectral resolution in the HI VGPS is 1.56 \kms with rms noise of 2K and in the \13co GRS the spectral resolution  is 0.21 \kms with rms of 2K.

%__________________________________________________________________

%%%%%% SECTION 3 - RESULTS %%%%%%%%%%%%%%%%%%%%%%%%%%
%%%%%% SECTION 3 - RESULTS %%%%%%%%%%%%%%%%%%%%%%%%%%

\section{Results}
\label{sec:results}

In this section we describe the characteristics of the spectral features and the morphology of the Scutum tangency traced by \cii and \niino.   We use the V$_{LSR}$ velocity structure in the line profiles to  infer that the spiral arm gas components are roughly distributed in different layers of the arm ranging from low density WIM to the dense molecular gas as  in the illustration of the distribution of ISM spiral arm gas components shown in Figure~\ref{fig:fig1}.

These layers display the gas evolutionary sequence  whereby the WIM falls into the gravitational potential, recombines, and is successively converted to higher density components, diffuse atomic, diffuse molecular, and dense molecular clouds, by either compression, cloud collisions, or convergent flows. The tangencies  are unique regions where the structural layers and dynamics of the arm can be more readily separated.  Furthermore, we use the velocity distribution to separate the WIM from the neutral layers. Thus by observing this region spectroscopically we can sample the different gas layers of the arm, and separate the  WIM's \nii and \cii emission from the other ionized components.
											
\subsection{\cii  and \nii spectra}
\label{sec:CII_NII_spectra}

\begin{figure*}[!ht]
 \centering
             \includegraphics[width=18cm]{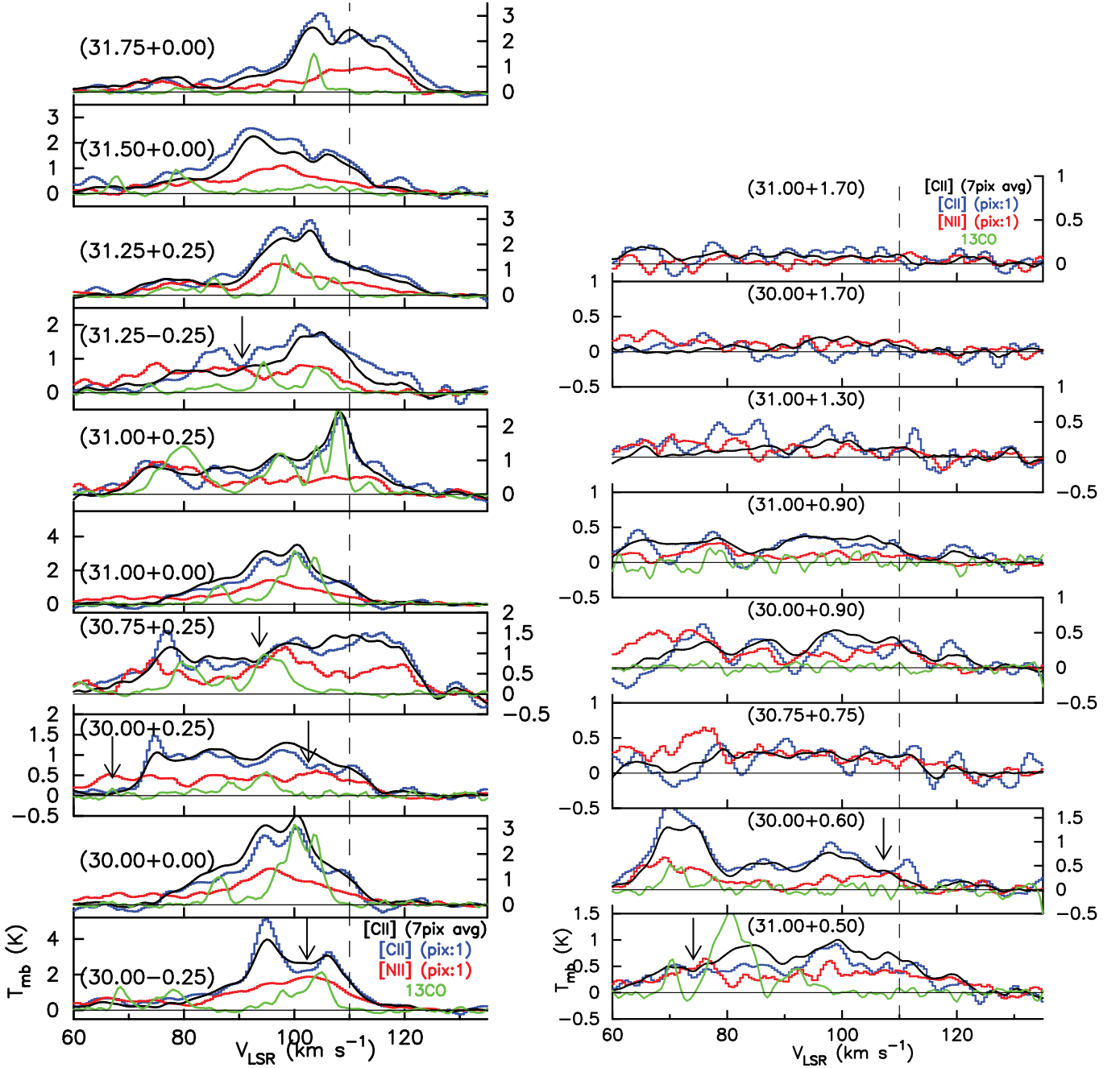}
      \caption{(left) The emission spectra of \cii (blue and black), \nii (red), and $^{13}$CO (green) for the ten lines of sight that are within $b= \pm$0\fdg25 are plotted as $T_{mb}$ versus velocity.  The central \cii pixel is plotted in blue and the average of the seven \cii pixels is plotted in black.  The single and average \cii emission profiles are similar but not exactly identical owing to variations in emission across the footprint of the upGREAT array. The black vertical dashed line marks the  tangent  velocity of the Scutum spiral arm.  (right) Figure caption the same as in the left panel for the eight lines of sight with $b > $0\fdg25.  The downward black arrows mark a few of the clearly evident  \cii absorption features.}
          \label{fig:fig5}
% \end{figure}
 \end{figure*}

In Figure~\ref{fig:fig5} (left panel) we plot the main beam temperature, $T_{mb}$, for \ciino, \niino, and $^{13}$CO(1-0)  versus $V_{LSR}$ for ten lines of sight across the Scutum arm ($V_{LSR} \sim$ 60 to 120 \kmsno), from longitude $l$ = 30\fdg0 to 31\fdg75 and for $b$ within $\pm$0\fdg25. In Figure~\ref{fig:fig5} (right panel) we plot these tracers for the eight lines of sight with $b >$ 0\fdg25.  For each LOS  the  red and blue lines show \nii and the central pixel of \ciino, respectively, while the black line shows the average of the 7 \cii pixels.  The green lines show the $^{13}$CO(1-0)  profile, which is used as a proxy for the velocity structure of the denser molecular spiral arm gas component. The \cii traces both neutral  and ionized gas components while \nii traces only the ionized gas.  The vertical dashed line indicates the tangent velocity as derived from assuming a Galactic rotation model \cite[e.g.][]{Roman-Duval2009}.  

In general the \cii central pixel and 7-pixel average are similar but there are small differences, as was also shown in the comparison of individual \cii pixels in Figure~\ref{fig:fig4}. The \cii central pixel (blue)  and \nii (red) profiles  represent emissions observed with comparable beams of 14\arcsec and 18\arcsec sizes, respectively, which allows a direct comparison of their intensities. On the other hand, the \cii 7 pixel average profiles (black) correspond to emission smoothed to a larger angular beam, $\sim$ 60\arcsec, comparable to the beam size of  the $^{13}$CO ($\sim$45\arcsecno) and \hi ($\sim$60\arcsecno) data. To display the velocity structure within the Scutum arm more clearly in Figure~\ref{fig:fig5}  we  plot only the velocity range of the Scutum arm, V$_{LSR}$= 60 -- 130 \kmsno.

Comparison of the intensities of the \ciino, $I$(\ciino), central pixel (blue) to \niino, $I$(\niino), (red) as a function of $V_{LSR}$ show regions where $I$(\ciino) to $I$(\niino), is less than 1.5, including some velocities where \nii is stronger than \ciino.  Analysis of the predicted contribution of \cii from \nii regions (see Section~\ref{sec:sec4.2}) shows that this ratio can not be less than 1.7 and more typically should be less than 2.  The low ratio seen at some velocities in Figure~\ref{fig:fig5} is a result of foreground and/or self-absorption.  We have indicated a few of these absorption regions with downward arrows.  Evidence for \cii absorption, sometimes strong absorption has been noted by \cite{Langer2016} towards other LOS observed in \nii and \cii with HIFI. 

% Figure 5
% Figure 5
% \begin{figure}

  \subsection{Longitudinal and latitudinal distribution}
 \label{sec:across_arm}

The \nii  and \cii emission for the spectra within  $|b| < $0\fdg25 covering $l$ = 30\fdg0 to 31\fdg75 in the Scutum tangency, as shown in Figure~\ref{fig:fig5},  extend over the same velocity range.  In contrast, $^{13}$CO is present only over portions of the velocity range of  the Scutum tangency.  The total emission is a blend of several cloud components as can be seen from the complexity in the $^{13}$CO $(l-v)$ map in Figure~\ref{fig:fig2}.  While the \nii distribution approximately follows that of \ciino, the ratio of  $T_{mb}$ of \cii to \nii varies with $V_{LSR}$.  In some lines of sight the \nii emission is nearly as strong as \ciino.  This association is seen in other studies of \nii and \cii and, as noted above,  is due in part to \cii absorption by low excitation foreground material \citep{Langer2015N}.

The latitudinal distribution of  the emission shown in Figure~\ref{fig:fig5} (right panel) reveals that $^{13}$CO is virtually absent for $b >$ 0\fdg6 but that \cii and \nii extend at least up to $b$ = 1\fdg3 and perhaps to 1\fdg7.  Closer to the plane, the emission from \nii and \cii extend across the entire velocity range of the Scutum tangency, while $^{13}$CO is restricted to a few velocity components. Taken together, the longitudinal and latitudinal distributions indicate that highly ionized gas, as traced by \nii and \ciino, and weakly ionized gas, as traced by a portion of \cii are present throughout all layers of the Scutum arm tangency.

  \subsection{Velocity structure of spiral arm gas components}
 \label{sec:across_arm}

% Figure 6
% Figure 6
 \begin{figure}
% \begin{figure*}[!ht]
 \centering
                    \includegraphics[width=9.20cm]{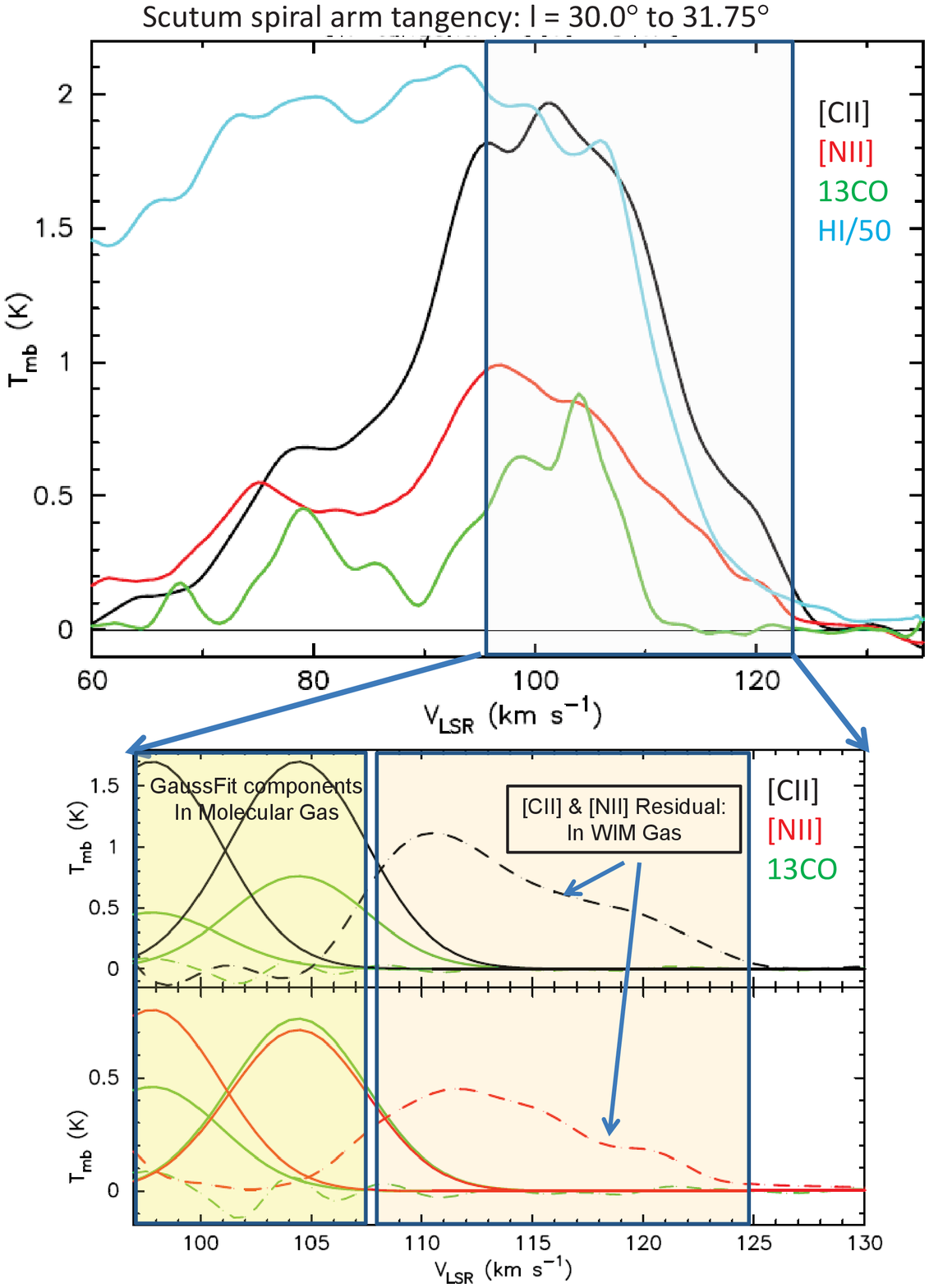}
      \caption{(top) The spectra for the atomic, molecular, and ionic gas tracers, \hi (blue), $^{13}$CO (green), \cii  (black), and \nii (red), averaged over all LOS within $b=\pm$0\fdg25 plotted as main beam temperature,
$T_{mb}$, versus $V_{LSR}$. The ionized gas tracers are seen to extend to much higher velocities, corresponding to the innermost leading edge of the Scutum arm, than the molecular gas as traced by $^{13}$CO.  (bottom) An expanded view of the ionized gas emission after subtracting out the \cii and \nii emission arising from the molecular clouds traced by $^{13}$CO.  The solid green curves represent two of the gaussian components fit to the $^{13}$CO emission.   The dashed green curve is the residual $^{13}$CO after subtraction.  Gaussian profiles were used to fit the \cii and \nii emission associated with the gas at velocities $<$ 110 \kmsno, while  keeping their $V_{LSR}$ at the peak and FWHM the same as the $^{13}$CO, and only adjusting their peak intensities.  Thus what remains for $V_{LSR} \ge$ 110 \kms is the \cii and \nii excess arising solely from the highly ionized gas with no associated CO.}
         \label{fig:fig6}
%        \end{figure*}
 \end{figure}

To characterize the overall distribution of gas across the Scutum arm we averaged the spectra from all longitudes ($l$=30\fdg0 to 31\fdg75) whose lines of sight fall within $b=\pm$0\fdg25.  This averaging improves the signal-to-noise of the weak WIM emission and  avoids biasing the results by observations towards any compact regions.   In Figure~\ref{fig:fig6} we plot the resulting $T_{mb}$ from the atomic, molecular, and ionized gas tracers versus $V_{LSR}$. The top panel in Figure~\ref{fig:fig6} compares $T_{mb}$ versus $V_{LSR}$ over the full velocity range of the Scutum tangency (note only the central upGREAT pixel is used so that \cii and \nii correspond to the same LOS).  The shaded rectangle ($V_{LSR}$ = 98 to 125 \kmsno) designates the emission from the inner Scutum arm where the different gas layers are more easily separated.  It is clear that the $^{13}$CO emission does not extend beyond $\sim$110 \kms while \cii and \nii extend up to at least 125 \kmsno.  Thus the \cii and \nii extend outside the  edge of the molecular gas tangency.  The \nii emission beyond 110 \kms likely arises primarily from the highly ionized gas at the inner edge of the Scutum arm.

To calculate the distribution and properties of the highly ionized gas we need to isolate the velocities over which \nii and \cii  characterize this gas alone versus the emission that comes from PDRs associated with the $^{13}$CO molecular gas. It has been shown by \cite{Velusamy2015} that the emissions in the gas components trace different gas layers which are well separated in  $V_{LSR}$ across the spiral arm tangency. To separate these components we made a multi-gaussian fit to the composite \niino, the \cii central pixel, and $^{13}$CO spectra in Figure~\ref{fig:fig6} across the velocity range 60 to 130 \kmsno.  The high end of this velocity range represents the WIM at the inner edge of the arm and it is highlighted  by the shaded box  in the upper panel of Figure~\ref{fig:fig6}. To separate out any \cii or \nii contribution in this box that is associated with the molecular gas traced by $^{13}$CO, we use the gaussian fits to $^{13}$CO to isolate the molecular components, as follows.   We fit the \cii and \nii emission by adjusting the peak intensity of a set of Gaussian profiles associated with the gas at velocities $<$ 110 \kmsno,   keeping their $V_{LSR}$ at the peak and FWHM the same as that for $^{13}$CO.  In the lower panel of Figure~\ref{fig:fig6} we show the two highest velocity $^{13}$CO gaussian components within the box.  We then subtracted out the respective gaussian components for \cii and \nii and plot the remaining excess in the right hand sub-panel, which represents just the \cii and \nii contributions from the highly ionized gas at the innermost edge of the Scutum arm.

 The strong correlation between the \cii and \nii emission in the right-subpanel is indicative of \cii emission arising from ionized gas with negligible contribution from neutral \hi gas (see Section~\ref{sec:sec4.2}).  The two shaded regions in the lower panel represent schematically the molecular gas (left) and WIM gas (right). The fact that the $^{13}$CO  residual (dashed green line) is negligible  shows that the emission in all gas components within the molecular gas layers of the Scutum spiral arm are well accounted for by the two Gaussian profiles (green lines) with linewidths (FWHM) $\sim$7 -- 8 \kms and centered at  $V_{LSR}$ $\sim$ 98 and 105 \kmsno, respectively. Therefore,  we conclude that the \cii and \nii residuals (black and red profiles, respectively) in the lower panel clearly delineate WIM gas that is well isolated from the molecular spiral arm. Note that the two Gaussian fits  to the molecular spiral arm are also consistent with $^{13}$CO in the ($l$--$V_{LSR}$) map shown in Figure~\ref{fig:fig2}.

%%%%%% SECTION 4 - DISCUSSION  %%%%%%%%%%%%%%%%%%%%%%%%%%
%%%%%% SECTION 4 - DISCUSSION  %%%%%%%%%%%%%%%%%%%%%%%%%%

\section{Discussion}
\label{sec:discussion}

Our SOFIA observations provide a rich data set of velocity resolved \cii and \nii emissions in the Scutum tangency. Our results show  widespread clear detections of \cii and \nii along all LOSs, especially, at velocities $V_{LSR}$ = 60 to 125 \kms and at $|b|$ $<$ 0\fdg25.  Although weaker at higher latitudes there is detectable emission at least up to $b$ = 1\fdg3.  The unique combination of \cii and \nii provides a powerful tool to analyze the ionized gas component in spiral arms and the ability to isolate \cii emission in the neutral and ionized gases.  

In general the WIM is a highly ionized, low density, $n$(e) $\sim$0.02 to 0.1 cm$^{-3}$, high temperature, $T_k$= 6000 to 10,000 K  gas that fills a 2 - 3 kpc layer around the Galactic midplane \cite[cf.][]{Ferriere2001,Cox2005,Haffner2009}. It is estimated to contain about 90\% of the ionized gas in the Galaxy and has a filling factor of order 0.2 to 0.4, which may depend on height above the plane.  At the midplane the filling factor may be as small as 10\%.  It has been studied with pulsar dispersion measurements \citep{Cordes2003}, H$\alpha$ emission \citep{Haffner2009,Haffner2016}, and nitrogen emission lines in the visible  \citep{Reynolds2001}.  More recently the WIM has been detected in  absorption with \nii  \citep{Persson2014}. The \cii and \nii emission lines can also be used to map  the WIM, however, the low densities of the WIM make it difficult to detect their emission throughout most of the Galaxy.  However, the spiral arm tangencies offer a unique geometry that enables the detection of the WIM in \nii and \cii because they correspond to regions with a very long path length in a relatively narrow velocity dispersion.  The WIM has been detected in \cii  emission by \cite{Velusamy2012,Velusamy2015} along the tangencies of the Perseus, Norma, Scutum, and Crux arms, and they derive higher densities than the interarm WIM. Clearly the WIM in spiral arms is different from that throughout the Galactic disk.  In this Section we investigate  the extended properties of the warm ionized medium as traced by \cii and \nii in the tangency region of the Scutum arm.

\begin{figure}
% \begin{figure*}[!ht]
 \centering
             \includegraphics[width=8.9cm]{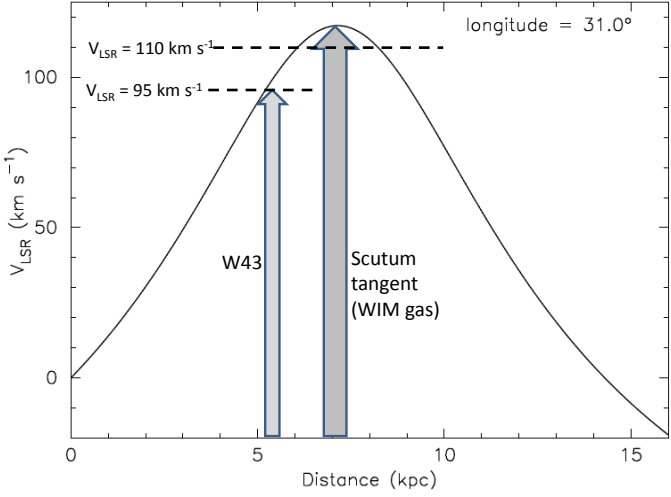}
      \caption{The distance--$V_{LSR}$ relationship for the Scutum arm tangency at $l$ = 31\fdg0 derived using the rotation curve of \cite{Reid2014}.  It shows that the WIM component of the Scutum tangency is from 6 -- 8 kpc, centered at $\sim$7 kpc distance, while the prominent star forming and \hii region, W43, is at about 5.5 kpc.}
      \label{fig:fig7}
       \end{figure}

For the analysis and interpretation of the \cii and \nii emission we adopt a distance of $\sim$7 kpc for the Scutum tangency estimated using the observed $V_{LSR}$ and the Galactic rotation model. In Figure~\ref{fig:fig7} we show the distance--$V_{LSR}$ relationship for the Galactic longitude of the Scutum tangency ($l$= 31\fdg0) derived using the rotation curve given by \cite{Reid2014} in their Figure 4 and the Galactic rotation parameters for their Model A5. The rotation velocities in this model for Galactocentric distance, $R_G >$ 4 kpc, is quite valid for the location of the Scutum tangency.  The WIM component observed at $V_{LSR}  >$ 110 \kms is located at the distance of the tangent points, $\sim$7 kpc and extends about $\pm$1 kpc.  We note that the distance--$V_{LSR}$ relationship plotted in Figure~\ref{fig:fig7} yields a distance of 5.5 kpc for the star forming region W43 (for $V_{LSR}$ = 95 --110 \kmsno) which is consistent with a distance of 5.5 kpc, obtained from maser parallaxes \citep{Zhang2014}.  Thus the Galactic bar-spiral arm interaction region containing W43 appears to be %on the 
just outside the outer part of the Scutum tangency (see Figure~\ref{fig:fig9}). 

% Figure 7
% Figure 7

%% Electron density from NII and CII
%% Electron density from NII and CII

\subsection{Electron density of the Scutum WIM}
 \label{sec:sec4.1}
 
 Nitrogen ions have two fine structure transitions, $^3$P$_1\rightarrow^3$P$_0$ at 205 \micron and $^3$P$_2\rightarrow^3$P$_1$ at 122 \micronno, and these can be used with an excitation model to calculate the electron density and column density $N$(N$^+$).  The 122 \micron line needs to be observed from space or a stratospheric balloon due to atmospheric opacity.  As we only have one \nii transition towards the Scutum tangency observed from SOFIA we cannot determine density $n$(e) and column density $N$(N$^+$) independently, however, we can get an idea of the range of $n$(e) using an approach outlined in \cite{Langer2015N} that depends on having an estimate of the size of the emission region.  For optically thin emission from a uniform region the intensity of the $^3$P$_1\rightarrow^3$P$_0$ transition can be written  \citep{Langer2015N},
 
\begin{equation}
I_{ion}({\rm [N\,II]})=0.156 x_{-4}({\rm N^+})L_{pc}n({\rm e})f_1(n({\rm e}),T_k) \,\,{\rm(K\,km\,s^{-1})\,},
\label{eqn:equation1}
\end{equation}

\noindent where $I_{ion}$(\niino) is the intensity in K \kmsno, $x_{-4}$(N$^+$) is the fractional abundance of N$^+$ in units of 10$^{-4}$, $L_{pc}$ is the size of the emission region in pc, T$_k$ is the kinetic temperature, and $f_1$(n(e),T$_k$) is the fractional population of the $^3$P$_1$ state, which is a function of $T_k$ and $n$(e). In the optically thin limit one can solve exactly for the fractional population of the levels as a function of $n$(e) for a given kinetic temperature \citep{Goldsmith2015}.  At the high kinetic temperatures associated with highly ionized gas, the kinetic energy of the electrons is orders of magnitude larger than the excitation energies required to populate the $^3$P$_1$ and $^2$P$_{3/2}$ states and the solutions are only weakly dependent on T$_k$ through the collisional rate coefficients.  For example for the collisional de-excitation rate coefficients calculated by \cite{Tayal2011} the temperature dependence varies from $\sim$ T$_k^{-0.3}$ to T$_k^{-0.5}$ depending on the transition.

Equation~\ref{eqn:equation1} can be solved iteratively for $n$(e) as a function of $I_{ion}$(\niino), using the balance equations for the population of the $^3P_2$ and $^3P_1$ \cite[see][]{Goldsmith2015}, assuming a reasonable value of T$_k$,  and given the size of the emission region and the fractional abundance of N$^+$.  The critical density (where the collisional de-excitation rate equals the radiative rate) for the $^3$P$_1$ level, $n_{cr}(e)$, is $\sim$175 cm$^{-3}$, so that at low densities, $n$(e)$<<n_{cr}$(e), Equation~\ref{eqn:equation1} can also be solved approximately, within $\pm$15\%, using,

\begin{equation}
n(e) = a_0 \Big[\frac{I(\rm{[N\,II]})}{x_{-4}(\rm{N^+})L_{pc}}\Big]^\alpha \,\,{\rm(cm^{-3})}\,,
\label{eqn:equation2}
\end{equation}

\noindent where, at a characteristic WIM temperature, T$_k$=8000 K,  $a_0$= 18.0 and $\alpha$= 0.51 for $n$(e)$\le$11 cm$^{-3}$, and $a_0$ =22 and $\alpha$= 0.72, for 11$<$$n$(e)$\le$ 100 cm$^{-3}$.  

 We consider the average properties of two Galactic zones, one within the plane and the other above the plane.  In the plane we average over all LOS within $b= \pm$0\fdg25, and above the plane we average all LOS within $b$ =0\fdg60 to 0\fdg90 (we exclude the LOS at $b$ = 1\fdg3 and 1\fdg7 because they are too noisy). By using all the data within these two regions we improve the signal to noise and minimize the effects of any localized sources.  We adopt a value for  the fractional abundance $x$(N$^+$) = 1.4$\times$10$^{-4}$ \cite[see Section 6.3.2][]{Goldsmith2015}, appropriate for the Galactic radius, $R_G$ = 4 kpc. From the geometry of the Scutum outer arm (see Figure~\ref{fig:fig9} and \cite{Reid2016}) the range $l \sim$ 30\deg to 32\deg  corresponds to a path length through the tangency $\sim$1$\pm$0.5 kpc, so we adopt $L$ =1 kpc to estimate $n$(e).  (In the low density case (see Equation~\ref{eqn:equation2}) $n$(e) is $\propto$ $L^{-0.5}$ so in choosing $L$=1 kpc the uncertainty in $n$(e) is of order $\pm$25\%.)  We solved for $n$(e) as a function of $I_{ion}$(\niino) over the velocity range  110 to 125 \kmsno.   We list the intensities and $<$$n$(e)$>$ in Table~\ref{tab:Table_1}. Typical average densities in the plane, are $<$$n$(e)$>$$\sim$0.9 cm$^{-3}$ and above the plane $<$$n$(e)$>$$\sim$ 0.4 cm$^{-3}$.

We can also calculate the density of the WIM from \ciino, providing that we have managed to isolate its WIM contribution from other sources (PDRs, CO-dark H$_2$, \hi clouds), using an approach similar to that derived for \niino.  For  $n$(e) less than the critical density for exciting the $^2$P$_{3/2}$ level of C$^+$, $n_{cr}$(e) $\sim$45 cm$^{-3}$, the electron density is given by \cite[see][]{Velusamy2012},

\begin{equation}
n(e)=2.92 T^{0.18}_k  \Big[\frac{I(\rm{[C\,II]})}{x_{-4}(\rm{C^+})L_{pc}}\Big]^{0.5} \,\,{\rm(cm^{-3})}\,.
\label{eqn:equation3}
\end{equation}

 \noindent   To solve for $n$(e) from \cii we adopt $T_k$=8000 K (Equation~\ref{eqn:equation3} is a very weak function of $T_k$) and $x$(C$^+$)= 2.9$\times$10$^{-4}$ \cite[see Equation 2 in][]{Pineda2013} appropriate to $R_G$ = 4 kpc.   We only use the central \cii pixel so that we are making a direct comparison of the \nii and \cii along the same LOS.  We find typical values  in the plane, $|b|\le$ 0\fdg25,  $<$$n$(e)$> \sim$0.8 cm$^{-3}$  and above the plane, $\sim$0.3 cm$^{-3}$, similar to those derived from \niino.
 
% This yields a carbon to nitrogen ratio $x$(C$^+$)=2.1$x$(N$^+$) appropriate to a Galactic radius of 4 kpc. 

% Figure 8
% Figure 8
% \begin{figure}
 \begin{figure*}[!ht]
 \centering
       \includegraphics[width=11cm,angle=-90]{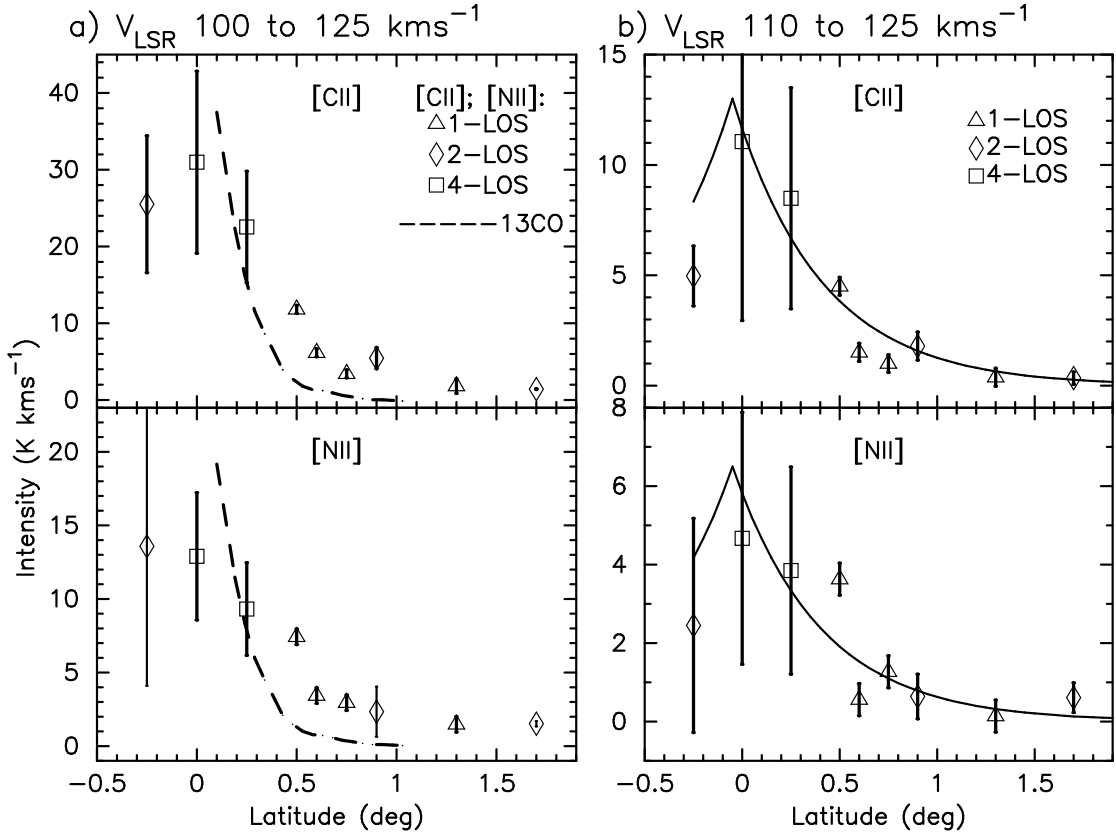}
      \caption{(left panels) The integrated intensities of \ciino, \niino, and $^{13}$CO (dashed line)  over the velocity range 100 to 125 \kms versus latitude. The symbols indicate how many SOFIA LOS were averaged to calculate the emission intensity.  The $^{13}$CO intensity profile is computed using the GRS data cube and we show only high latitude values to demonstrate how sharply it drops to nearly zero by $b$ = 0\fdg5.  $^{13}$CO  has a much smaller scale height than \cii and \niino. (right panels)  The integrated intensities of \cii and \nii over the velocity range 110 to 125 \kms versus latitude. There is no measurable $^{13}$CO in this velocity range, and the \cii and \nii arise solely from the WIM.  The exponential curves correspond to exp(-$(b-\delta b)/b_0$ with $\delta b$ = -0\fdg05 and $b_0$ = 0\fdg45. }
         \label{fig:fig8}
% \end{figure}
 \end{figure*}

 \subsection{\cii from the Scutum WIM}
 \label{sec:sec4.2}

As shown in \cite{Langer2016} we can calculate how much  \cii emission comes from the highly ionized gas, $I_{ion}$(\ciino), in the optically thin limit from the intensity of \niino, using the following relationship,

\begin{equation}
I_{ion}{\rm([C II])} =0.675 \frac{f_{3/2}({\rm C^+})}{f_1({\rm N^+})} \frac{x{\rm(C^+)}}{x{\rm(N^+)}} I_{ion}{\rm(N II)} \,\,,
\label{eqn:equation4}
\end{equation}

\noindent where  $f_i$ is the fractional population of the corresponding levels.  In Equation~\ref{eqn:equation4} $f_{3/2}/f_1$ is only weakly dependent on density for $n$(e)=10$^{-3}$ to 10$^2$ cm$^{-3}$, and ranges from 1.22 to 1.73\footnote{At low densities, $n$(e)$\le$5 cm$^{-3}$, $f_{3/2}/f_1$ ranges from 1.53 to 1.73 and to a good approximation we can set this ratio to a constant of 1.63 to within $\pm$6\%. The fractional abundance ratio, $x$(C$^+$)/$x$(N$^+$) = 2.1 yields a simple estimate  for the \cii emission arising from the highly ionized gas traced by \niino, $I_{ion}$(\ciino) =2 .3$I_{ion}$(\niino), when $n$(e)$\le$ 5 cm$^{-3}$.}, and we adopt $x$(C$^+$)/$x$(N$^+$) = 2.1\footnote{We adopt a solar fractional abundance of $x$(N$^+$) = 6.76$\times$10$^{-5}$ \citep{Asplund2009} with an abundance gradient -0.07 dex kpc$^{-1}$  \citep{Shaver1983}, and  for C$^+$, $x$(C$^+$)= 1.4$\times$10$^{-4}$, with the same gradient as $x$(N$^+$) \citep{Pineda2013}, which yields a ratio, $x$(C$^+$)/$x$(N$^+$) = 2.1.}.

In Table~\ref{tab:Table_1} we give the \cii intensity calculated to arise from the WIM, $I_{ion}$(\ciino), corresponding to $V_{LSR}$ =110 -- 125 \kmsno, and the fraction of the total observed intensity, $I_{tot}$(\ciino) (the molecular regions are discussed in Section~\ref{sec:HighDensity}).  The \cii intensity calculated to arise from the highly ionized gas, $I_{ion}$(\ciino), from just the WIM, $V_{LSR}$ =110 -- 125 \kmsno, ranges from 2.0 K \kms out of the plane ($b$=0\fdg6 to 0\fdg9) to 9.7 K \kms in the plane ($b$=$\pm$0\fdg25).    In and out of the plane the average fraction $I_{ion}$(\ciino)/$I_{tot}$(\ciino) arising from the WIM is $\sim$ 90\%. Thus the \cii emission from the WIM is a significant fraction of the observed  \cii intensity towards the WIM, as might be expected from this highly ionized low density medium with very little neutral gas. 

%\begin{equation}
%I_{ion}{\rm([C II])} =2.4 I_{ion}{\rm(N II)}.
%\label{eqn:equation5}
%\end{equation}

 If $\sim$90\% of the \cii emission comes from the highly ionized WIM, then the remaining $\sim$10\% probably comes from the \hi gas clouds mixed in with the WIM, as there is no molecular gas present.  In Figure~\ref{fig:fig6} we see that there is \hi emission from 110 to 125 \kmsno.  In Table~\ref{tab:Table_1} we give the average $I$(\hino) in and above the plane, which can readily be converted to an equivalent column density, $N$(C$^+$)=1.82$\times$10$^{18}$$x$(C$^+$)$I$(\hino).  The corresponding C$^+$ column densities in and above the plane are, $N$(C$^+$) = 2.6 and 1.3 $\times$10$^{17}$ cm$^{-2}$, respectively.  This column density corresponds to a mean hydrogen density, $<$$n$(H)$>$ = 1.2 and 0.6 cm$^{-3}$, respectively.  However, because the \hi clouds do not fill the LOS, their local density is likely to be  much higher.   For purposes of estimating their contribution to $I$(\ciino) we assume a typical kinetic temperature $T_{k}$ = 100 K and typical \hi cloud densities $n$(H) = 10 -- 20 cm$^{-3}$. The \hi contribution to $I$(\ciino), $I_{{\rm HI}}$(\ciino), even for the larger value of $n$(H)= 20 cm$^{-3}$, is $\sim$ 4\% in the plane and $\sim$10\% out of the plane.  This percentage contribution is consistent with the difference between total \cii intensity and the $\sim$90\% contribution of the WIM to the total \cii intensity.
																				
 \subsection{z-distribution}
 \label{sec:z-distribution}

In principle the scale height of the gas in the disk is a measure of the different forces in the disk, such as gravity and pressure, and thus the energy insertion in the disk.  The mass distribution arises from stars, the ISM, and dark matter.  The pressure is a result of thermal energy, random motion of clouds, cosmic ray and magnetic pressure, while supernovae and OB-associations provide energy injection that may drive gas into the halo.   \cii observations with {\it Herschel} HIFI made possible the determination of the mean distribution of the gas traced by C$^+$.  However, because \cii traces many different gas components, to get the scale heights of the different ISM components it is necessary to separate out the different scale heights by associating the emission coming from CO, \hino, and WIM regions.  \cite{Velusamy2014} analyzed the distribution of all the \cii emission detected by the GOT C+ survey as a function of $z$ and found that  the average scale height for \cii (all gas components) is $\sim$ 170 pc, larger than that for CO, but smaller than that for \hino, as might be expected. They found that \cii from the WIM and \hi combined had a scale height $\sim$330 pc, larger than \hi alone.  However, with \cii alone, they were unable to separate the scale height of the weakly and highly ionized gas.  From other studies the highly ionized diffuse gas is known to have a number of components and extend high above the plane.

To determine the $z$--distribution of the ionized gas traced by \cii and \nii we averaged the spectra for each value of $b$ using observations at adjacent longitudes.  This approach reduces the risk that we might have an unusually bright source of emission in the beam and improves the signal-to-noise at the higher latitudes where the emission is weak.  To derive the $z$-distribution of the intensity of \cii and \nii in the WIM we restrict the calculation of intensity to $V_{lsr}$ = 110 to 125 \kms where there is no CO gas (see Figure~\ref{fig:fig6}).

In Figure~\ref{fig:fig8} we plot the integrated intensity of \cii and \nii as a function of $b$, for two different velocity ranges, 100 -- 125 \kms (left panel) and 110 -- 125 \kms (right panel), where the symbols indicate the number of averaged LOS.  The left panel also includes the integrated $^{13}$CO intensity profile (dashed line) as a function of $b$.  The $^{13}$CO intensity profile is computed using the GRS data cube \citep{Jackson2006}. Note that the CO profile near  $b$ = 0\fdg0 is uncertain due to complexity in the distribution of strong $^{13}$CO emission as seen in Figure~\ref{fig:fig2}. Nevertheless the high latitude values  demonstrate how sharply the \13co emission  drops. It is nearly zero by $b$ = 0\fdg5, and evidently has a much smaller scale height than \cii and \niino. In the right panel, where the velocity range corresponds only to WIM gas without any evident molecular component, we find that the data points are roughly  consistent with  an exponential, $I$=$I_0$exp(-$(b-\delta b)/b_0$) with   a scale height  $b_0$ = 0\fdg45  and offset from plane $\delta b$ = 0\fdg05. Within the uncertainty in the measured intensities both \cii and \nii have similar $z$-scales of $\sim$55 pc assuming a distance of 7 kpc to the tangency.  These scale heights  do not measure the distribution of the gas density but are measures of the \cii and \nii emission from the WIM, because their intensities are proportional to $n$(e)$^2$, and thus at some point below the sensitvity of the \cii and \nii surveys. 
 
\cii observations with {\it Herschel} HIFI made possible the determination of the radial distribution of the gas traced by C$^+$.   However, because \cii traces many different gas components it is necessary to separate out the different scale heights by determining the C$^+$ emission coming from CO, \hino, and WIM regions.  To separate out the different \cii sources, \cite{Velusamy2014} correlated the intensities in 3 \kms wide velocity bins, ``spaxels", in each \cii velocity profile with the corresponding spaxel intensities in the $^{12}$CO,  $^{13}$CO and \hi velocity profiles,  and found that the \cii gaussian scale height for $z$ in the Scutum tangency is smaller than the average value for \cii for the disk as a whole.  There are a number of possible reasons for the difference. The spaxel analysis was a global average across the disk which had more sensitivity at higher values of $b$ and thus may detect more emission higher in the plane.  Second, the Scutum tangency has many active regions of star formation and thus potentially more heating UV.  Thus the \cii emission in the plane ($|b| \lesssim$ 0\fdg5) may be much stronger than the radial average in the plane found by \cite{Pineda2013} and thus skew the fit to a smaller scale height.

% Figure 9
% Figure 9
\begin{figure}
% \begin{figure*}[!ht]
 \centering
                          \includegraphics[width=8.9cm,angle=0]{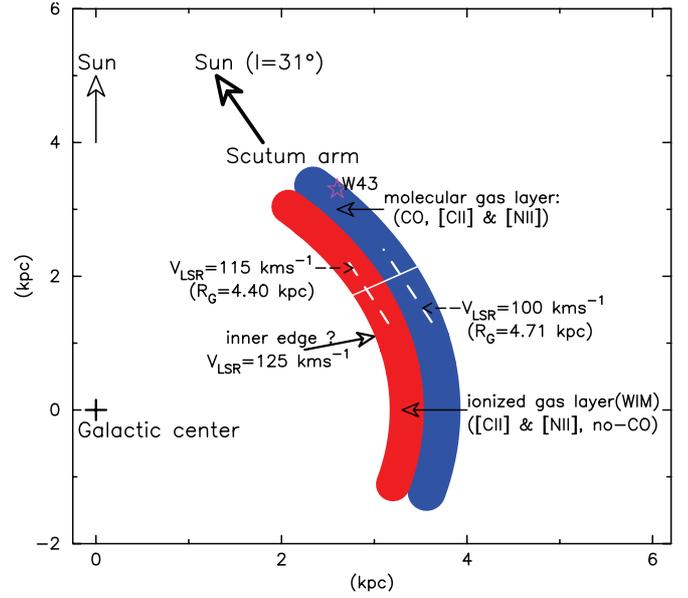}
      \caption{A schematic radial cross section view of the spiral arm perpendicular to the LOS to Scutum spiral arm tangency showing the inner to outer edges of the spiral arm (this sketch is not to scale). The molecular gas layer traced by $^{13}$CO is shown by a thick blue line. Dashed lines show two representative radial velocities and corresponding Galactic locations.   The red arc indicates the spiral density wave shock. The location of W43, from the study by \cite{Zhang2014}, is marked on the schematic.}
         \label{fig:fig9}
 \end{figure}
% \end{figure*}

 \subsection{The Source of \nii and \cii in the Scutum arm}
\label{sec:models}
The schematic diagram in Figure~\ref{fig:fig9} illustrates the proposed structure of the gas layers of the Scutum arm as inferred from the observed  velocity structure of the gas components traced by \ciino, \niino, and $^{13}$CO. In this picture only the WIM is traced by \nii and \cii at $V_{LSR} >$  110 \kms and the red curve indicates where a shock might form, while the denser molecular gas layer is traced by $^{13}$CO at  $V_{LSR} <$   110  \kmsno.  We use the unique viewing geometry (as shown in the Figure~\ref{fig:fig9}) of the tangency to derive the cross-section view across the spiral arm, from the inner to outer edge, as a function Galacto-centric radial distance (R$_G$). We use the $V_{LSR}$--R$_G$ relationship assuming pure Galactic rotation, with rotation velocities of 235 \kms and 220 \kms at R$_G$ =4 kpc and R$_{sun}$ = 8.34 kpc, respectively \cite[c.f.][]{Reid2014,Reid2016}.  Thus  the velocity resolved \ciino, \niino, and $^{13}$CO spectral data delineate the gas layers of the spiral arm.  The low surface brightness \cii and \nii emission observed at $V_{LSR} >$  110 \kmsno, which are without associated $^{13}$CO, is clearly from low density ionized  WIM gas located along the inner boundary  of the spiral arm.  On the other hand the brighter \cii and \nii emission seen with  associated $^{13}$CO at  $V_{LSR} <$   110  \kms is clearly from the denser ionized gas mixed with  the population of dense molecular gas along the mid--layers of the spiral arm. Using the \cii emission alone to trace ionized gas, in combination with CO, \cite{Velusamy2015} have shown   in the Norma and Perseus tangencies similar evidence for the segregation of the WIM ionized gas layer being displaced inward towards the Galactic center with respect to the dense molecular gas layers of the spiral arm.  In the present case of the Scutum tangency, the combination of both \cii and \nii tracing the ionized gas component  supports  our prior interpretation of the gas layers distributed from inner to outer  edges of the spiral arm as shown in the schematic in Figure~\ref{fig:fig9}.

\nii is widespread and relatively strong throughout the Scutum spiral arm tangency with typical intensities $I_{ion}$(\niino) $\sim$ (0.37 -- 0.53) $I$(\ciino) (Table~\ref{tab:Table_1} column 7).  This ratio is much larger than $\sim$ 0.1 to 0.2 found by COBE FIRAS to be typical of the Galactic distribution.  The SOFIA $b$ scans show that \nii is strongest at the midplane, where the dense molecular cloud tracer $^{13}$CO is strongest, and decreases with latitude exponentially with a scale height of 0\fdg57 and is essentially undetectable beyond 1\fdg7. Thus  the \nii emission observed by COBE FIRAS, with a 7\deg beam,  would have been significantly beam diluted. Here we discuss the conditions and sources of the relatively large electron densities responsible for the \nii emission and the large \nii to \cii ratio in the Scutum arm and by extension in the  sparse galactic surveys of \nii and \cii \citep{Goldsmith2015,Langer2016}.

\subsubsection{WIM component}

The electron densities in the WIM derived from \cii emission  in \cite{Velusamy2012,Velusamy2015}  and from \nii in this paper, are much larger than those characteristic of the Galactic interarm WIM derived using other probes. In the velocity range $V_{LSR}$ = 110 -- 125 \kmsno, where there is no evidence of molecular clouds as indicated by the absence of $^{13}$CO,  the \nii emission arises almost entirely from the WIM.  This result is supported by our estimate that  $\sim$90\% of the observed \cii  intensity is associated with the N$^+$ gas in this velocity range.  The WIM average electron density  within $b$ = $\pm$0\fdg25  is $\sim$0.4 to 0.9  cm$^{-3}$, or about 5 to 20 times larger  than the average WIM density, 0.03--0.08 cm$^{-3}$ \citep{Haffner2009}, that fills the bulk of the Galactic disk.  We interpret this density increase in the inner edge of the arm as due to compression of the interarm WIM in the gravitational potential of the leading edge of the inner spiral arm.

\begin{table*}[!htbp]																		
\caption{Properties of the Scutum arm ionized gas} 
\label{tab:Table_1}															
\begin{tabular}{ccccccccccc}
\hline	
$b$ range &  $V_{LSR}$& $I_{ion}$(\niino)  &   $<$$n$(e)$>$ & $I$(\ciino) & $<$$n$(e)$>$ &  $I$(\niino)/ & $I_{ion}$(\ciino) & $I_{ion}$(\ciino)/ & $I$(\hino)& $I_{\rm{HI}}$(\ciino)$^a$\\
  averaged &  km/s &K km/s & cm$^{-3}$    & K km/s &  cm$^{-3}$ &  I(\ciino) & K km/s & $I_{tot}$(\ciino) & K km/s & K km/s \\
  \hline
  \hline
%-0\fdg25  to +0\fdg25 &110--125 & 3.66 & 0.9$^b$ &  9.92 &  0.7$^b$ &  2.7 & 8.8  & 0.86 & 495  & 0.2 \\
-0\fdg25  to +0\fdg25 &110--125 & 4.05 & 0.9$^b$ &  10.8 &  0.8$^b$ &  0.37 & 9.7  & 0.89 & 521  & 0.5 \\
0\fdg60 to 0\fdg90 &  110--125 & 0.82 & 0.4$^b$ &  2.2 & 0.3$^b$ &  0.38  & 2.0 & 0.92 & 251 & 0.2 \\
-0\fdg25  to +0\fdg25 & 70--110 & 25.9 & 40$^c$ &  49.2 &  18$^c$ &  0.53 & 44.9  & $<$0.91$^d$ & 3700  & 3.4 \\
\hline	
\end{tabular}
\\	
a) The \cii intensity from the \hi gas is calculated assuming $n$(H) = 20 cm$^{-3}$. b)  $n$(e) for the WIM is derived from $I_{ion}$(\niino) and $I$(\ciino)\\ assuming $L$ = 1 kpc.   c) $n$(e) in the ionized boundary layers is derived from $I_{ion}$(\niino) and $I$(\ciino) assuming $L$ = 8 pc. d) This value is only an upper limit as there is evidence for foreground and/or self-absorption of \cii -- see text.
 \end{table*}												
%\end{table}

\subsubsection{High density ionized gas in the Scutum molecular layer}
\label{sec:HighDensity}

Unlike the WIM component observed at $V_{LSR} >$ 110 \kms  interpreting the strong \nii and \cii emission features at
$V_{LSR}<$ 110 \kms is more challenging as they are mixed with the denser molecular gas component in the spiral arm.  These LOSs are located  in the W43 complex (W43-south) (see Figure~\ref{fig:fig9}), which shows evidence for shocks and compression flows \cite[e.g.][]{Carlhoff2013} with enhanced star formation.  To analyze individual features, we ideally need the densities and/or physical size of the emitting regions.  In principle, if we could observe spectrally resolved \nii at 122 \micron and 205 \micron we could constrain the density of the ionized gas in the emitting region as a function of $V_{LSR}$.  However, the high spectral resolution HIFI instrument on  {\it Herschel} operated only beyond 157 \micron and  {\it Herschel} PACS only had a resolving power of $\simeq$1000 at 122 \micronno.  Presently no heterodyne instrument is available to observe the 122 \micron line.

In their Galactic survey \cite{Goldsmith2015} observed \nii with PACS at  two lines of sight in the Scutum arm, ($l$,$b$)=(30\fdg0,0\fdg0) and (31\fdg2766,0\fdg0).  They used the two \nii transitions with an excitation model to  derive the corresponding electron densities, $n$(e) = 29 and 31 cm$^{-3}$ and column densities, $N$(N$^+$) = 7.7$\times$10$^{16}$ and 11.4 $\times$10$^{16}$ cm$^{-2}$.  They also observed  (31\fdg2766,0\fdg0)  in \nii 205 \micron with {\it Herschel} HIFI and the spectrum shows that nearly all the \nii emission was coming from the Scutum arm \cite[Figure 1 in ][]{Langer2016}.  The \nii spectrum of (30\fdg0,0\fdg0) included here (Figure~\ref{fig:fig5}) also shows \nii emission throughout the arm.  Furthermore, PACS detected \nii emission across the 25-pixel array which has a $\sim$45\arcsec footprint, corresponding to  about 1.5 pc  at the distance to the Scutum arm. Thus it is reasonable to surmise that the Scutum arm is an extended  source of dense highly ionized gas that coexists with the molecular gas, perhaps as an ionized boundary layer associated with the molecular clouds.

We can estimate the average path length of the emitting region observed with PACS, $L$=$N$(N$^+$)$x$(N$^+$)/$n$(e), by substituting the values for $n$(e) and $N$(N$^+$) from Table 2 in \cite{Goldsmith2015}.  We find $L$= 6.1 pc and 8.5 pc for the emitting regions in (30\fdg0,0\fdg0) and (31\fdg2766,0\fdg0), respectively.  However, because the PACS data are spectrally unresolved we cannot separate how much \nii 122 \micron emission comes from the WIM tangency and how much from within the molecular arm, which may include dense compact or extended \hii regions or shocked gas.  Thus an exact calculation of the density of the different ionized components from an excitation model is not possible without the spectrally resolved 122 \micron line.  

Instead, to estimate the average conditions of highly ionized gas in the molecular layers of the arm, we use only the emission in the velocity range where $^{13}$CO is detected, $V_{LSR}$ = 70 to 110 \kmsno.  In Table~\ref{tab:Table_1} (row 3) we list the \cii and \nii intensities for this velocity range.  If we assume that most of this emission comes from a region with a total length of 6 -- 8 pc, as discussed above,  then $n$(e) $\sim$ 40 cm$^{-3}$ from $I_{ion}$(\niino) and $\sim$18 from $I$(\ciino). Thus the densities derived here bracket the values in \cite{Goldsmith2015} and are consistent with the PACS excitation analysis. Such high densities are indicative of \hii regions and the ionized boundary layers (IBL) of dense clouds \citep{Abel2005,Abel2006}. The multiple velocity features seen in each of the HIFI and GREAT \nii spectra show that the total path length of 8 pc will then correspond to about 6 to 10  \hii or IBL components with an average individual depth of $\sim$1 pc.   . 

Table~\ref{tab:Table_1} (row 3) lists the intensity of \cii that arises from the \nii component, $I_{ion}$(\ciino) = 44.9 K \kmsno, derived using Equation~\ref{eqn:equation4}.  The predicted fractional \cii intensity from the highly ionized gas, $I_{ion}$(\ciino)/$I_{tot}$(\ciino) = 91\%.  In addition, as seen in Table~\ref{tab:Table_1}, the contribution of \cii from \hi clouds in the plane is  $\sim$7\%. This result is surprising because it would indicate that very little \cii emission comes from PDRs and CO-dark H$_2$ clouds, which have significant amounts of C$^+$ and much higher densities than the \hi clouds. However, as seen in Figure~\ref{fig:fig5} and discussed in Section~\ref{sec:CII_NII_spectra}, there is evidence for multiple \cii absorption features and therefore the observed $I_{tot}$(\ciino) is a lower limit to the intrinsic source intensity.    \cite{Langer2016} found similarly low \cii to \nii ratios, sometimes with \nii $\ge$ \ciino,  towards several of the ten LOS they observed in \cii and \nii with HIFI. In several components  they were able to correct for the absorption and  the percentage of emission from the ionized gas ranged from 34\% to 58\% instead of nearly 100\%.   Unfortunately the crowding of features along the LOS to the Scutum tangency, as seen in Figures~\ref{fig:fig2} and \ref{fig:fig5}, makes it difficult to correct for the \cii absorption using the approach described by \cite{Langer2016}.  The large fraction of \cii arising from \nii regions is not just a phenomenon in the Galactic plane as   \cite{Rollig2016} observed \cii and \nii with GREAT in the nearby spiral galaxy IC 342, and  showed that 30\% -- 90\% of \cii arises from the highly ionized regions traced by \niino. 

That the \nii emission is detected throughout the velocity range of the molecular gas, $V_{LSR}$ = 70 to 110 \kmsno, indicates that the ionized boundary layers are distributed throughout the arm.  There are two potential sources of the dense highly ionized gas, photoionization by O stars and shock compression of the WIM.  In the GRS survey, which includes the Scutum tangency, \cite{Anderson2009} report  high, $\sim$ 80\%, morphological association between CO and \hii regions, which could account for the wide spread \nii associated with $^{13}$CO in the SOFIA spectra.  However, a more likely source of the widespread dense $n$(e) in the \nii layers  is shock compression as the WIM gas falling into the gravitational potential of the arm, which can accelerate the WIM at the inner gas layer of the spiral arm to a velocity of several \kmsno, much greater than the sound speed in the neutral gas clouds, 
$\sim$1 \kmsno, resulting in a shock at the interface. This mechanism would also explain the high $n$(e) densities distributed along about 100 LOS throughout the inner galaxy  \citep{Goldsmith2015}. In the Scutum arm we might expect to see some shocked gas with enhanced \cii and \nii emissions due to the proximity of the end of the Galactic bar and spiral arm interaction region. However, the WIM (at $V_{LSR} >$ 110 \kmsno) is fully devoid of CO, that is without any molecular gas,  which is more consistent with compression of the interarm WIM by gravitational infall rather than gas streaming from the Galactic bar.   Therefore, in our model the ionized boundary layers are continually fueled by the incoming stream of gas.

%%%%%% SECTION 5 - SUMMARY %%%%%%%%%%%%%%%%%%%%%%%%%%
%%%%%% SECTION 5 - SUMMARY %%%%%%%%%%%%%%%%%%%%%%%%%%

\section{Summary}
\label{sec:summary}

We report high spectral resolution observations of \cii and \nii in longitude and latitude across the Scutum spiral arm tangency using upGREAT and GREAT, respectively,  on SOFIA.  We use these tracers to characterize the highly ionized gas of the WIM in terms of its electron density, the fraction of \cii that arises from the highly ionized gas, and the scale height of \cii and \nii emission.  The high spectral resolution observations allow us to separate the \nii and \cii emission from the WIM from that in the neutral arm. The WIM emission over the inner leading edge of the Scutum arm extends from $V_{LSR}$ = 110 to 125 \kmsno.  An analysis of the \nii  intensity from the WIM indicates an average electron density, $n$(e)$\sim$ 0.9 cm$^{-3}$ within $b$ = $\pm$0\fdg25, approximately one to two orders of magnitude denser than the interarm WIM.  This difference is indicative of compression of the interarm  WIM as it flows onto the Scutum arm.  The results of combining \nii and \cii support unambiguously the earlier results of \cite{Velusamy2012,Velusamy2015} on the properties of the WIM along spiral arm tangencies based on \cii alone.  We find that the scale height for the \cii and \nii emission,  $\sim$55 pc, extends beyond that of the molecular gas tracer $^{13}$CO.  A comparison of the \cii and \nii emission in the WIM shows that a significant fraction,  $\sim$90\% of the \cii arises from the highly ionized gas.

There is widespread, relatively strong \nii emission associated with the molecular regions of the Scutum arm, implying that there is highly ionized gas within the arm with $n$(e) $\sim$ 20 -- 40 cm$^{-3}$  in several thin clumps or layers,  totaling about 8 pc in depth.  This high density $n$(e) and accompanying \nii could be associated with \hii regions distributed throughout the arm, but is most likely the result of shock compression of the WIM,  accelerated by the gravitational potential of the Scutum arm. 
This process may also explain the strong \nii emission, high electron densities, and relatively large \nii to \cii emission observed in recent sparse \nii surveys of the inner Galaxy \citep{Goldsmith2015,Langer2016}. In the molecular arm we can only set an upper limit on the fraction of \cii emission that arises from the highly ionized gas traced by \nii because of absorption of the \cii emission by foreground gas.  More extensive maps of spectrally resolved \nii and \cii of the spiral arms are needed to examine the relatively strong \nii emission in the Scutum arm and along many other lines of sight in the Galactic disk.

% format apostrophe 's on Bill's mac 's's's's

%__________________________________________________________________

%%%%%% ACKNOWLEDGEMENT %%%%%%%%%%%%%%%%%%%%%%%%%%
%%%%%% ACKNOWLEDGEMENT %%%%%%%%%%%%%%%%%%%%%%%%%%

\begin{acknowledgements}
We thank the referee for a careful reading of the manuscript and a number of comments that improved the content. We also thank the SOFIA engineering and operations teams for their support which enabled the observations presented here.  The research reported here is based largely on observations made with the NASA/DLR Stratospheric Observatory for Infrared Astronomy (SOFIA).  SOFIA is jointly operated by the Universities Space Research Association, Inc. (USRA), under NASA contract NAS2-97001, and the Deutsches SOFIA Institut (DSI) under DLR contract 50 OK 0901. This work was performed at the Jet Propulsion Laboratory, California Institute of Technology, under contract with the National Aeronautics and Space Administration.   %{\copyright}2017 California Institute of Technology. 
USA Government sponsorship acknowledged.

\end{acknowledgements}

%__________________________________________________________________

%%%%%% REFERENCES %%%%%%%%%%%%%%%%%%%%%%%%%%
%%%%%% REFERENCES %%%%%%%%%%%%%%%%%%%%%%%%%%

%% the following two command lines are needed for using bibtex
\bibliographystyle{aa}
\bibliography{aa-2017-31198-refs}

\begin{thebibliography}{32}
\expandafter\ifx\csname natexlab\endcsname\relax\def\natexlab#1{#1}\fi

\bibitem[{{Abel}(2006)}]{Abel2006}
{Abel}, N.~P. 2006, \mnras, 368, 1949

\bibitem[{{Abel} {et~al.}(2005){Abel}, {Ferland}, {Shaw}, \& {van
  Hoof}}]{Abel2005}
{Abel}, N.~P., {Ferland}, G.~J., {Shaw}, G., \& {van Hoof}, P.~A.~M. 2005,
  \apjs, 161, 65

\bibitem[{{Anderson} {et~al.}(2009){Anderson}, {Bania}, {Jackson}, {Clemens},
  {Heyer}, {Simon}, {Shah}, \& {Rathborne}}]{Anderson2009}
{Anderson}, L.~D., {Bania}, T.~M., {Jackson}, J.~M., {et~al.} 2009, \apjs, 181,
  255

\bibitem[{{Asplund} {et~al.}(2009){Asplund}, {Grevesse}, {Sauval}, \&
  {Scott}}]{Asplund2009}
{Asplund}, M., {Grevesse}, N., {Sauval}, A.~J., \& {Scott}, P. 2009, \araa, 47,
  481

\bibitem[{{Carlhoff} {et~al.}(2013){Carlhoff}, {Nguyen Luong}, {Schilke},
  {Motte}, {Schneider}, {Beuther}, {Bontemps}, {Heitsch}, {Hill}, {Kramer},
  {Ossenkopf}, {Schuller}, {Simon}, \& {Wyrowski}}]{Carlhoff2013}
{Carlhoff}, P., {Nguyen Luong}, Q., {Schilke}, P., {et~al.} 2013, \aap, 560,
  A24

\bibitem[{{Cordes} \& {Lazio}(2003)}]{Cordes2003}
{Cordes}, J.~M. \& {Lazio}, T.~J.~W. 2003, ArXiv Astrophysics e-prints

\bibitem[{{Cox}(2005)}]{Cox2005}
{Cox}, D.~P. 2005, \araa, 43, 337

\bibitem[{{Ferri{\`e}re}(2001)}]{Ferriere2001}
{Ferri{\`e}re}, K.~M. 2001, Reviews of Modern Physics, 73, 1031

\bibitem[{{Goldsmith} {et~al.}(2015){Goldsmith}, {Y{\i}ld{\i}z}, {Langer}, \&
  {Pineda}}]{Goldsmith2015}
{Goldsmith}, P.~F., {Y{\i}ld{\i}z}, U.~A., {Langer}, W.~D., \& {Pineda}, J.~L.
  2015, \apj, 814, 133

\bibitem[{{Haffner} {et~al.}(2009){Haffner}, {Dettmar}, {Beckman}, {Wood},
  {Slavin}, {Giammanco}, {Madsen}, {Zurita}, \& {Reynolds}}]{Haffner2009}
{Haffner}, L.~M., {Dettmar}, R.-J., {Beckman}, J.~E., {et~al.} 2009, Reviews of
  Modern Physics, 81, 969

\bibitem[{{Haffner} {et~al.}(2016){Haffner}, {Reynolds}, {Babler}, {Madsen},
  {Hill}, {Barger}, {Jaehnig}, {Mierkiewicz}, {Percival}, {Chopra}, {Pingel},
  {Reese}, {Gostisha}, \& {Wunderlin}}]{Haffner2016}
{Haffner}, L.~M., {Reynolds}, R.~J., {Babler}, B.~L., {et~al.} 2016, in
  American Astronomical Society Meeting Abstracts, Vol. 227, American
  Astronomical Society Meeting Abstracts, 347.17

\bibitem[{{Heyminck} {et~al.}(2012){Heyminck}, {Graf}, {G{\"u}sten}, {Stutzki},
  {H{\"u}bers}, \& {Hartogh}}]{Heyminck2012}
{Heyminck}, S., {Graf}, U.~U., {G{\"u}sten}, R., {et~al.} 2012, \aap, 542, L1

\bibitem[{{Jackson} {et~al.}(2006){Jackson}, {Rathborne}, {Shah}, {Simon},
  {Bania}, {Clemens}, {Chambers}, {Johnson}, {Dormody}, {Lavoie}, \&
  {Heyer}}]{Jackson2006}
{Jackson}, J.~M., {Rathborne}, J.~M., {Shah}, R.~Y., {et~al.} 2006, \apjs, 163,
  145

\bibitem[{{Langer} {et~al.}(2016){Langer}, {Goldsmith}, \&
  {Pineda}}]{Langer2016}
{Langer}, W.~D., {Goldsmith}, P.~F., \& {Pineda}, J.~L. 2016, \aap, 590, A43

\bibitem[{{Langer} {et~al.}(2015){Langer}, {Goldsmith}, {Pineda}, {Velusamy},
  {Requena-Torres}, \& {Wiesemeyer}}]{Langer2015N}
{Langer}, W.~D., {Goldsmith}, P.~F., {Pineda}, J.~L., {et~al.} 2015, \aap, 576,
  A1

\bibitem[{{Persson} {et~al.}(2014){Persson}, {Gerin}, {Mookerjea}, {Black},
  {Olberg}, {Goicoechea}, {Hassel}, {Falgarone}, {Levrier}, {Menten}, \&
  {Pety}}]{Persson2014}
{Persson}, C.~M., {Gerin}, M., {Mookerjea}, B., {et~al.} 2014, \aap, 568, A37

\bibitem[{{Pineda} {et~al.}(2013){Pineda}, {Langer}, {Velusamy}, \&
  {Goldsmith}}]{Pineda2013}
{Pineda}, J.~L., {Langer}, W.~D., {Velusamy}, T., \& {Goldsmith}, P.~F. 2013,
  \aap, 554, A103

\bibitem[{{Reid} {et~al.}(2016){Reid}, {Dame}, {Menten}, \&
  {Brunthaler}}]{Reid2016}
{Reid}, M.~J., {Dame}, T.~M., {Menten}, K.~M., \& {Brunthaler}, A. 2016, \apj,
  823, 77

\bibitem[{{Reid} {et~al.}(2014){Reid}, {Menten}, {Brunthaler}, {Zheng}, {Dame},
  {Xu}, {Wu}, {Zhang}, {Sanna}, {Sato}, {Hachisuka}, {Choi}, {Immer},
  {Moscadelli}, {Rygl}, \& {Bartkiewicz}}]{Reid2014}
{Reid}, M.~J., {Menten}, K.~M., {Brunthaler}, A., {et~al.} 2014, \apj, 783, 130

\bibitem[{{Reynolds} {et~al.}(2001){Reynolds}, {Sterling}, {Haffner}, \&
  {Tufte}}]{Reynolds2001}
{Reynolds}, R.~J., {Sterling}, N.~C., {Haffner}, L.~M., \& {Tufte}, S.~L. 2001,
  \apjl, 548, L221

\bibitem[{{Risacher} {et~al.}(2016){Risacher}, {G{\"u}sten}, {Stutzki},
  {H{\"u}bers}, {Bell}, {Buchbender}, {B{\"u}chel}, {Csengeri}, {Graf},
  {Heyminck}, {Higgins}, {Honingh}, {Jacobs}, {Klein}, {Okada}, {Parikka},
  {P{\"u}tz}, {Reyes}, {Ricken}, {Riquelme}, {Simon}, \&
  {Wiesemeyer}}]{Risacher2016}
{Risacher}, C., {G{\"u}sten}, R., {Stutzki}, J., {et~al.} 2016, \aap, 595, A34

\bibitem[{{R{\"o}llig} {et~al.}(2016){R{\"o}llig}, {Simon}, {G{\"u}sten},
  {Stutzki}, {Israel}, \& {Jacobs}}]{Rollig2016}
{R{\"o}llig}, M., {Simon}, R., {G{\"u}sten}, R., {et~al.} 2016, \aap, 591, A33

\bibitem[{{Roman-Duval} {et~al.}(2009){Roman-Duval}, {Jackson}, {Heyer},
  {Johnson}, {Rathborne}, {Shah}, \& {Simon}}]{Roman-Duval2009}
{Roman-Duval}, J., {Jackson}, J.~M., {Heyer}, M., {et~al.} 2009, \apj, 699,
  1153

\bibitem[{{Shaver} {et~al.}(1983){Shaver}, {McGee}, {Newton}, {Danks}, \&
  {Pottasch}}]{Shaver1983}
{Shaver}, P.~A., {McGee}, R.~X., {Newton}, L.~M., {Danks}, A.~C., \&
  {Pottasch}, S.~R. 1983, \mnras, 204, 53

\bibitem[{{Stil} {et~al.}(2006){Stil}, {Taylor}, {Dickey}, {Kavars}, {Martin},
  {Rothwell}, {Boothroyd}, {Lockman}, \& {McClure-Griffiths}}]{Stil2006}
{Stil}, J.~M., {Taylor}, A.~R., {Dickey}, J.~M., {et~al.} 2006, \aj, 132, 1158

\bibitem[{{Tayal}(2011)}]{Tayal2011}
{Tayal}, S.~S. 2011, \apjs, 195, 12

\bibitem[{{Vall{\'e}e}(2008)}]{Vallee2008a}
{Vall{\'e}e}, J.~P. 2008, \apj, 681, 303

\bibitem[{{Velusamy} \& {Langer}(2014)}]{Velusamy2014}
{Velusamy}, T. \& {Langer}, W.~D. 2014, \aap, 572, A45

\bibitem[{{Velusamy} {et~al.}(2015){Velusamy}, {Langer}, {Goldsmith}, \&
  {Pineda}}]{Velusamy2015}
{Velusamy}, T., {Langer}, W.~D., {Goldsmith}, P.~F., \& {Pineda}, J.~L. 2015,
  \aap, 578, A135

\bibitem[{{Velusamy} {et~al.}(2012){Velusamy}, {Langer}, {Pineda}, \&
  {Goldsmith}}]{Velusamy2012}
{Velusamy}, T., {Langer}, W.~D., {Pineda}, J.~L., \& {Goldsmith}, P.~F. 2012,
  \aap, 541, L10

\bibitem[{{Young} {et~al.}(2012){Young}, {Becklin}, {Marcum}, {Roellig}, {De
  Buizer}, {Herter}, {G{\"u}sten}, {Dunham}, {Temi}, {Andersson}, {Backman},
  {Burgdorf}, {Caroff}, {Casey}, {Davidson}, {Erickson}, {Gehrz}, {Harper},
  {Harvey}, {Helton}, {Horner}, {Howard}, {Klein}, {Krabbe}, {McLean}, {Meyer},
  {Miles}, {Morris}, {Reach}, {Rho}, {Richter}, {Roeser}, {Sandell}, {Sankrit},
  {Savage}, {Smith}, {Shuping}, {Vacca}, {Vaillancourt}, {Wolf}, \&
  {Zinnecker}}]{Young2012}
{Young}, E.~T., {Becklin}, E.~E., {Marcum}, P.~M., {et~al.} 2012, \apjl, 749,
  L17

\bibitem[{{Zhang} {et~al.}(2014){Zhang}, {Moscadelli}, {Sato}, {Reid},
  {Menten}, {Zheng}, {Brunthaler}, {Dame}, {Xu}, \& {Immer}}]{Zhang2014}
{Zhang}, B., {Moscadelli}, L., {Sato}, M., {et~al.} 2014, \apj, 781, 89

\end{thebibliography}
%\bibliographystyle{apj}
%\bibliographystyle{aasjournal.bst}
 %-----------------------------------------------------------------------------------------------------------------------------------------------------------------------

\end{document}